\documentclass[11pt,twoside]{article}
\usepackage{asp2014}

\aspSuppressVolSlug
\resetcounters

\begin{document}

\title{Virtual Reality and Immersive Collaborative Environments: the New Frontier for Big Data Visualisation}

\author{Alexander~K.~Sivitilli$^1$, Angus~Comrie$^2$, Lucia~Marchetti$^{1}$, and Thomas~H.~Jarrett$^1$}
\affil{$^1$Department of Astronomy, University of Cape Town, Private Bag X3, Rondebosch 7701, Cape Town, South Africa; \email{sivi@ast.uct.ac.za}}
\affil{$^2$Inter-University Institute for Data Intensive Astronomy, Cape Town, South Africa}

\paperauthor{Alexander~K.~Sivitilli}{alexandersivitilli@gmail.com}{https://orcid.org/0000-0002-2563-4793}{University of Cape Town}{Astronomy Dept}{Cape Town}{Western Cape}{7701}{South Africa}
\paperauthor{Angus~Comrie}{accomrie@gmail.com}{0000-0002-1790-0705}{University of Cape Town}{Astronomy Dept}{Cape Town}{Western Cape}{7701}{South Africa}
\paperauthor{Lucia~Marchetti}{marchetti@ast.uct.ac.za}{https://orcid.org/0000-0003-3948-7621}{University of Cape Town}{Astronomy Dept}{Cape Town}{Western Cape}{7701}{South Africa}
\paperauthor{Thomas~H.~Jarrett}{tjarrett007@gmail.com}{0000-0002-4939-734X}{University of Cape Town}{Astronomy Dept}{Rondebosch}{Western Cape}{7701}{South Africa}



  
\begin{abstract}
The IDIA Visualisation Laboratory based at the University of Cape Town is exploring the use of virtual reality technology to visualise and analyse astronomical data. The iDaVIE software suite currently under development reads from both volumetric data cubes and sparse multi-dimensional catalogs, rendering them in a room-scale immersive environment that allows the user to intuitively view, navigate around and interact with features in three dimensions.

This paper will highlight how the software imports from common astronomy data formats and processes the information for loading into the Unity game engine. It will also describe what tools are currently available to the user and the various performance optimisations made for seamless use. Applications by astronomers will be reviewed in addition to the features we plan to include in future releases.

\end{abstract}

\section{Introduction}

With big datasets becoming more common in the next generation of astronomical observations and simulations, there is an increasing need to refine our methods in how we process and analyze them. Consequently, it is critical to re-evaluate the tools and techniques used in reducing data to discernible forms compatible with our senses as humans. It is with this aspiration that the IDIA Visualisation Lab (IVL) at the University of Cape Town was established as a joint project between UCT Astronomy and the \textit{Inter-University Institute for Data Intensive Astronomy} (IDIA). The primary goal of the IVL is to explore new ways of interacting with and analyzing big datasets. Alongside other state-of-the-art immersive technologies \citep{O5-5_adassxxviii}, a major component of this project is the \textit{Immersive Data Visualization Interactive Explorer} (iDaVIE) virtual reality (VR) software suite currently under development.

The aim of iDaVIE is to render datasets in a roomscale 3D space where users can intuitively view and uniquely interact with their data in ways unafforded by conventional flatscreen and 2D solutions. Viewing includes both the conversion of datasets of typical machine-readable formats to practical representations along with the ability for the user to navigate spaces of the virtual setting to see the data from multiple translational, rotational and scalable viewpoints. Unique interaction with the data in VR entails modifying viewing parameters, taking measurements, and annotating in the same space where the data is rendered, effectively allowing the user to remain in the VR session and perform science directly on the data.

\section{Development Tools}

We develop iDaVIE through the Unity game engine \citep{unitywebsite} using the SteamVR plugin \citep{steamvrwebsite} that enables the flexible use of VR headsets from multiple manufactures. Current hardware platforms that have been successfully tested include Oculus Rift and Rift S, HTC Vive and Vive Pro, as well as the Samsung Odyssey. Beta release (see Section \ref{future}) will be distributed as a packaged binary.

\section{Data Import and Rendering} \label{import}

The iDaVIE software suite currently consists of two separate solutions, iDaVIE-v and iDaVIE-c, respectively focusing on the rendering of volumetric data cubes and sparse multi-dimensional catalogs.

With volumetric data cubes in the form of three-dimensional binary FITS images, iDaVIE-v parses the file with the cfitsio library \citep{cfitsiowebsite} and stores the data as a pointer in memory. This is  handled using the C++ native plugin capabilities of Unity. A 3D texture object is then created in Unity. For each pixel in each eye's viewport, a ray marching shader is run, tracing a ray from the user's view through the 3D texture, and accumulating data values. Values can be excluded from the accumulation based on value, location in the 3D texture, or from a mask lookup in a separate 3D texture. After accumulation, a color map is applied.

For catalogs, iDaVIE-c has the ability to parse text-based tables in IPAC-table format as well as binary FITS tables. Minimum parameters required are x,y,z spatial coordinates that are mapped to the table with a configurable JSON file. The file may also include the source labels from other columns of the table to be used for different rendering parameters such as color map, opacity, particle size, and particle shape. The required columns of the table are then stored in buffers on the GPU. A GPU shader pipeline transforms these data points to rendered images. For each data point, a vertex shader transforms the point's location to the correct 3D position, based on the user's camera position and the chosen scale of the dataset. A geometry shader then turns this location into a camera-facing quadrilateral (a technique known as billboarding), with a width determined from the the mapping configuration and the relevant column entry of the data point. Adjustments to the standard billboarding approach are made to ensure that quadrilaterals are rotated for each eye's position and orientation. Finally, a pixel shader performs texture lookup, color mapping and opacity adjustments based on the mapping configuration and the relevant columns of the data point. 

\section{Interaction}
In both software solutions, the VR user is able to navigate by either physically moving around the virtual environment (and in fact, the room within which you stand) or by using navigation gestures with the controllers, which can be more conducive to interaction and long sessions. The gesture navigation capabilities include translation, rotation, as well as scaling of the rendered dataset. Currently the data files must be selected in the Unity editor prior to launching the VR session. There are also specific interactions afforded to the user depending on the data type.

With volumetric data, the user can crop the rendered cube to show a subset that, if previously down-sampled (see Section \ref{performance}), may be displayed in higher resolution. The user accomplishes this by drawing a box around the desired subset of the data using a 3D cursor (hand controller), and using a voice command to confirm the crop. These boxes can also be drawn in the scene to indicate features for external observers and saved as an exported table of dataset annotations in a text-based format. The exported table can then be reloaded with rendered boxes at a later session. A mask can also be applied to the cube in addition to adjusting bounds and style of the applied color map. Most of the interaction with the volumetric data can be done through the use of voice commands recognised by the microphone built into the VR headset.

For catalogs, iDaVIE-c allows adjusting the applied color map directly by the VR user, however adjustment of other rendering parameters in real-time requires the assistance of an operator at the Unity desktop editor interface. The operator is also able to manually adjust rendering parameters listed in Section \ref{import} as well as load and/or swap datasets for the user. 

\section{Performance} \label{performance}

\articlefigure{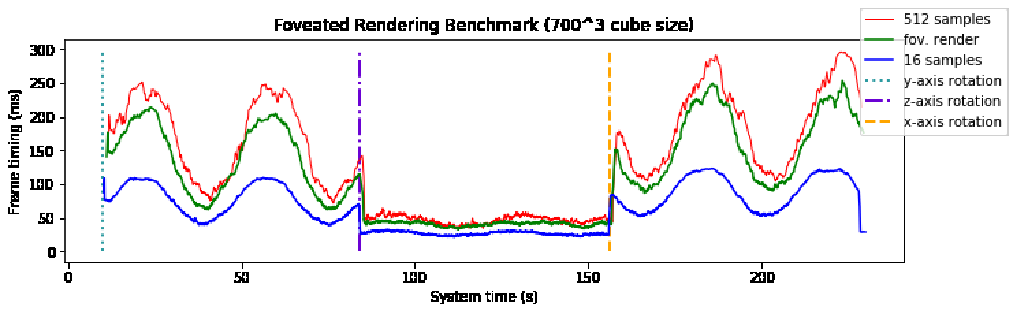}{fig:plot}{Benchmark results for iDaVIE-v from a randomly generated cube. The cube is rotated around all three axes with start times indicated by vertical lines. Maximum ray-marching sampling is shown in the top plot (red), foveated rendering the middle (green), minimum ray-marching sampling is on the bottom (blue).  Note that lower frame timing means better performance.}

With large datasets it is inevitable that we need to push the performance of the software to its limits, as well as the hardware on which it is running. However, low framerates and delayed interactions at levels considered merely inconvenient in
traditional 2D software solutions can actually make VR interaction unusable. User experience in VR requires a minimum ideal framerate to prevent nausea and discomfort \citep{Weech2019}. It is with this consideration that we adopt target frame-timings for the VR above which we must consider the performance of iDaVIE. 

Ideal performance in iDaVIE-c is accomplished by capping the number of data points for importing tables, and tasking an operator at the desktop editor to manually adjust particle size and opacity to prevent cases of overdraw.

With iDaVIE-v, we need to be mindful of the size of the volumetric cubes being imported. There are a few emerging limitations relevant to the voxel pipeline. As we currently load the entire cube into memory in the parsing process, enough system RAM is needed to accommodate the  float array. The 3D texture object is then created from a down-sampled version of the cube, reduced using a multi-threaded block averaging process. The target cube size keeps both the individual dimensions of the texture under 2048 voxels (software limit of Unity) and the overall texture under a predefined maximum size, locking the spatial axes' downscale factors in the process. We also experiment with reducing sample size of the ray-marching algorithm, both globally and dependent on projected distance from the left/right center of the VR screens, effectively harnessing a custom fixed foveated rendering scheme. We hope to adapt this scheme to a dynamic form when eye tracking in VR headsets becomes readily available.
 
Although initial performance evaluation was done using qualitative feedback of the user and monitoring frame timing of the VR system,  we are currently in the process of standardizing how we benchmark the two software solutions (see figure \ref{fig:plot}). This involves loading a randomly generated volumetric cube or data cloud, rotating along the three axes at fixed distances from the VR camera, and comparing the resulting frame timings recorded from the SteamVR software. It is worthwhile to note the drastic difference in frame timings in the figure depending on cube orientation. This is a consequence of the 3D textures in Unity using one-dimensional data structures to store the voxel information, favoring the x-axis and severely crippling framerates when looking down the z-axis. We are currently experimenting
with using different memory and i/o storage schemes to improve performance.

\section{Applications and Future}  \label{future}

iDaVIE is currently undergoing pilot use cases over a range of data types from astronomical observations to theoretical simulations, to  cellular and microbiology 3D scans \citep{O5-5_adassxxviii}.
May 2020 will see the the official release of iDaVIE-v 1.0, the beta version of the volumetric data software. This will also be the deadline for including a new round of features including a desktop GUI (developed in collaboration with the \textit{Istituto Nazionale di Astrofisica} (INAF) in Catania), the ability to edit voxels of a mask, realtime histogram plotting, and a state-saving feature.
We also plan to add a few essential features in the long-term to both solutions. This includes collaboration tools for other users, both VR and external, to interact with the VR session, either locally or remotely; cloud database interfacing; motion and time-domain representation; greater data format flexibility to expand application outside of astronomy; and digital planetarium integration.

\bibliography{P4-6}


\end{document}